\documentclass{article}
\usepackage{preamble}
\title{A Wilson Line Realisation of Quantum Groups}

\author{Nanna Aamand$^1$ \and Dani Kaufman$^1$ }

\date{%
    $^1$\begin{small}
        University of Copenhagen \\
         Department of Mathematical Sciences \\
        2100 Copenhagen, Denmark \\[2ex]
    \end{small} 
    \today
}
\begin{document}
\maketitle

\begin{abstract}
    We study Wilson line operators in 3-dimensional Chern-Simons theory on a manifold with boundaries and prove to leading order, through a direct calculation of Feynman integrals, that the merging of parallel Wilson lines reproduces the coproduct on the quantum group $U_\hbar(\mathfrak g)$. We outline a connection of this theory with the moduli spaces of local systems defined by Goncharov and Shen.
\end{abstract}
\section{Introduction}

Topological quantum field theories like Chern-Simons theory have long been known to have connections to quantum groups and their representations. The goal of this paper is to give an explicit realization of this connection in the setting of perturbative 3-dimensional Chern-Simons theory with boundary conditions, by showing that the tensor product in the category of representations of the quantum group can be realized from the operation of merging two parallel Wilson lines.

In recent papers \cite{integrabilityI}, \cite{integrabilityII} Costello, Witten and Yamazaki constructed a 4-dimensional conformal Chern-Simons theory in which they realized the representation theory of the Yangian. They suggest in section 7.8 of their second paper that a 3-dimensional theory with boundaries can be constructed from their 4-dimensional theory by restricting to $U(1)$-invariant fields. The corresponding 3-dimensional theory was constructed explicitly by the first author in \cite{Aamand:2019evs}. Concretely, the set-up
is Chern-Simons theory on a manifold $M = \R^2 \times I$ with a semi-simple Lie algebra $\g$. The relevant set of boundary conditions on the gauge field $A\in \Omega^1(M,\g)$ comes from defining a Manin triple $(\mathfrak l_-,\mathfrak l_+,\g)$ and restricting $A$ to take value in subalgebras $\mathfrak l_+$ (resp. $\mathfrak l_-$) on the upper (resp. lower) boundary. Since the gauge symmetry of the action is broken by the boundary conditions, this theory permits a set of gauge invariant operators given by open Wilson lines associated to representations of $\g$ and extending to infinity along $\R^2$. 

In the present paper, we consider a product on the set of Wilson lines in the above setting coming from merging two parallel lines. By computing the leading order Feynman amplitude for a gauge boson coupling to the pair of merging lines, we show that this product agrees with the leading order deformation of the tensor product in $\Rep(U_\hbar(\g))$. 
It was argued in \cite{Aamand:2019evs} that, in the same theory, the leading order contribution to the expectation value of a pair of crossing Wilson lines is given by the classical $r$-matrix. Together these results suggest that the category of Wilson line operators is equivalent to the category $\Rep (U_\hbar(\g))$ as a braided monoidal category.

One motivation for considering this 3d Chern-Simons theory with boundary conditions instead of the 4d conformal version is its close connection to the moduli spaces of local systems on punctured surfaces considered by Goncharov and Shen, \cite{goncharov2022quantum}. In fact, as we will discuss in the final section of this paper, 
our construction can be seen as a realization through perturbation theory of the ``geometric avatar of a TQFT'' described in section 5 of the paper of Goncharov and Shen.
\section{The Gauge Theory}\label{sec:CS-theory}
The gauge theory that we will be concerned with is 3-dimensional Chern-Simons theory defined by the action:
\begin{equation}\label{CS-action}
    S_\text{CS}(A)=\frac{1}{2\pi}\int_{M}\Big<A\wedge dA+\frac{1}{3}A\wedge [A, A]\Big>
\end{equation}
where the gauge field (connection) $A\in\Omega^1(M,\g)$ is a 1-form on $M$ taking values in the Lie algebra $\g$ of the gauge group and $\braket{\,,\,}$ is an invariant symmetric bilinear form on $\g$. In the present paper we will take $M=\R^2\times I$ where $I=[-1,1]$ and we take $\g$ to be a complex semi-simple Lie algebra.

In order to have a well-defined theory in the presence of boundaries, we must impose boundary conditions on the gauge field. Specifically, when varying the action with respect to the gauge field, $A\to A+\delta A$ where $\delta A$ is an exact 1-form, we pick up a boundary term:
\begin{align}\label{boundary term}
    \frac{1}{2\pi}\int_{\R^2\times \{-1,1\}} \left<A\wedge \delta A\right>,  
\end{align}
and we must impose boundary conditions on $A$ and $\delta A$ ensuring that this term vanishes on each boundary. At the same we want that the restriction of the gauge theory to each boundary component is in itself a well-defined, gauge invariant theory (see \cite{integrabilityI} section 9.1 for more elaboration on this). By the second requirement, choosing a set of boundary conditions amounts to specifying subalgebras $\mathfrak l_+,\mathfrak l_-\subset \g$ and imposing that $A$ and $\delta A$ take value in $\mathfrak l_+$ (resp. $\mathfrak l_-$) at the upper (resp. lower) boundary. It was argued in \cite{integrabilityI} that a we get a valid set of boundary conditions giving rise to quantum group structures by choosing $\mathfrak l_+$ and $\mathfrak l_-$ so that the triple $(\g,\mathfrak l_+,\mathfrak l_-)$ is a Manin triple. In other words $\mathfrak l_+$ and $\mathfrak l_-$ must be non-intersecting, half-dimensional, isotropic subalgebras of $\g$ such that $\g=\mathfrak l_+\oplus \mathfrak l_-$.

\section{Manin Triples and Quantum Groups}\label{sec:Lie}
\subsection{Constructing a Manin Triple}\label{sec:Manin}
Not all semi-simple Lie algebras admit the structure of a Manin triple (for example if the Lie algebra has odd dimension). Following the construction of \cite{integrabilityI} (section 9.2) we can modify $\g$ to accommodate for this by adding another copy of the Cartan subalgebra. We give here the construction in full detail.
 
Let $\h\subset \g$ be a Cartan subalgebra 
and consider the root system $\Phi$ of $\mathfrak g$ relative to $\h$ equipped with a polarization $\Phi=\Phi_+\sqcup\Phi_-$. 
We write $\mathfrak n_+$ and $\mathfrak n_-$ for the sum over root spaces $\g_\alpha$ corresponding to the set of positive roots $\alpha\in \Phi_+$ and negative roots $-\alpha\in\Phi_-$, respectively. Then $\mathfrak n_-$ and $\mathfrak n_+$ are isotropic subalgebras and we get a decomposition of $\g$:
$$
\g = \mathfrak n_-\oplus \h \oplus \mathfrak n_+.
$$
Now add to $\g$ another copy $\tilde \h$ of the Cartan subalgebra:
$$\tilde \g=\g\oplus\tilde\h$$ 
with the bracket on $\g$ trivially extended to $\tilde \g$, i.e. $[a,\tilde b]=0$ for $a\in \g$ and $\tilde b \in \tilde\h$.
We can extend the Killing form on $\g$ to $\tilde \g$ as follows: $\braket{a,\tilde b}=0$ for all $a\in\g$, $\tilde b\in  \tilde \h$ and $\braket{\tilde a,\tilde b}=\braket{a,b}$ for all $a,b\in\h$. This gives an invariant symmetric bilinear form on $\tilde \g$. Define
\begin{equation*}
\begin{aligned}
    &\h_+=\big\{h+ \mathbf{i}\tilde h\,\big|\,h\in\h\big\}\subset \h\oplus\tilde \h 
    \\
    &\h_-=\big\{h-\mathbf{i}\tilde h\,\big|\,h\in\h\big\}\subset \h\oplus\tilde \h\,. 
\end{aligned}
\end{equation*}
With this definition $\braket{\,,\,}$ vanishes on $\h_+$ and $\h_-$ and thereby choosing
\begin{equation}\label{eq:bc}
        \mathfrak l_-=\mathfrak n_-\oplus\mathfrak h_- \ \text{ and } \
        \mathfrak l_+=\mathfrak n_+\oplus \h_+.
\end{equation}
the triple $(\tilde \g, \mathfrak l_+, \mathfrak l_-)$ is a Manin triple.

\paragraph{Conventions} 
Let $r$ be the rank of $\g$. We fix a choice of simple roots $\Delta_+ = \{\alpha_i\}_{i=1}^r \subset \Phi_+$ along with a basis $\{H_i\}_{i=1}^r$ of $\mathfrak h$. Furthermore, we
let $X_\alpha$ be a generator of the root space $\g_\alpha$ normalized so that $[X_\alpha,X_\beta] = X_{\alpha+\beta}$ and $[X_{\alpha_i},X_{-\alpha_i}] = H_i$. Using standard notation, we write $E_i, F_i \coloneqq X_{\alpha_i}, X_{-\alpha_i}$ for each simple root $\alpha_i\in $ $\Delta_+$ and we write $H_i^+=H_i+\mathbf{i}\tilde H_i$ and $H_i^-=H_i-\mathbf{i}\tilde H_i$. A basis $\mathcal B_+$ for $\mathfrak l_+$ can now be given as:
$$
\mathcal B_+=\{X_\alpha,H_i^+\}_{\alpha\in \Phi_+,i=1,\dots, r}.
$$ 
Let $\mathcal B_-$ be the basis for $\mathfrak l_-$ dual to $\mathcal B_+$ with respect to the Killing form. Then $\mathcal B=\mathcal B_+\cup \mathcal B_-$ is a basis for $\tilde \g$. We write $$\mathcal B = \{t_a\}_{a=1,\dots, \dim \tilde \g}\ , \ \ \mathcal B_+ = \{t_a\}_{a=1,\dots,\dim \tilde \g/2}\,.$$ Finally, we denote by $t^a$ the dual element of $t_a\in \mathcal B$. 
\subsection{Quantization}\label{sec:quantization}
In this subsection we briefly recall the construction of a quantum double 
via the Drinfel'd double construction. For a detailed exposition we refer the reader to e.g. \cite{Gautam:2022xpq} section 4.

Let $\mathfrak b=\mathfrak n_+\oplus \mathfrak h\subset \g$ be the Borel subalgebra relative the to setup of section \ref{sec:Manin} and let $(a_{ij})$ be the Cartan matrix of $\g$. The Drinfel'd double $\mathfrak D_\hbar(\mathfrak b)$
is the algebra over $\mathbb C[[\hbar]]$ with generators:
$$\{E_i,F_i,H^+_i,H^-_i~|~i=1,\dots,r\}$$ and relations 
\begin{equation}
\begin{aligned}\label{eq:brackets}
&[H_i^\pm,E_j]=a_{ij}E_j   
\\
&[H_i^\pm,F_j]=-a_{ij}F_j
\end{aligned} \hspace{1.5cm}
\begin{aligned}
&[H^\pm_i,H^\pm_j]=[H^\pm_i,H^\mp_j]=0
\\
&[E_i,F_j]= \delta_{ij}\frac{e^{\hbar H_i^+/2}-e^{-\hbar H^-_i/2}}{e^{\hbar/2}-e^{-\hbar/2 }}
\end{aligned}
\end{equation}
along with the quantum Serre relations for $i\neq j$. In the case of $\mathfrak{sl}_n(\mathbb C)$ these relations take the form 
\begin{equation*}
\begin{aligned}
    E_i^2E_{j}-(e^{\hbar/2}+e^{-\hbar/2 })E_iE_{j}E_i + E_iE_{j}^2 = 0\\
    F_i^2F_{j}-(e^{\hbar/2}+e^{-\hbar/2 })F_iF_{j}F_i + F_iF_{j}^2 =0.
\end{aligned}
\end{equation*}
For the general case see e.g. \cite{Gautam:2022xpq} section 4.2. The quantized universal enveloping algebra of $\g$ is constructed from the double as 
$$
U_\hbar(\g)\coloneqq \mathfrak D_\hbar(\mathfrak b)/\braket{H_i^+-H_i^-}.
$$
It holds that $\mathfrak D_\hbar(\mathfrak b)$ has the structure of a quasi-triangular Hopf algebra with co-product:
\begin{equation}\label{eq:generator_coprod}
\begin{aligned}
&\Delta E_i=1\otimes E_i +E_i\otimes 1+\frac{\hbar}{4}(E_i\otimes H^+_i-H^+_i\otimes E_i)+\mathcal O(\hbar^2) \\
&\Delta F_i=1\otimes F_i +F_i\otimes 1 +\frac{\hbar}{4}(F_i\otimes H^-_i-H^-_i\otimes F_i) + \mathcal O(\hbar^2)\\
&\Delta H^\pm_i= 1\otimes H^\pm_i +H^\pm_i\otimes 1 .
\end{aligned}
\end{equation}
Notice that this realizes the usual co-product on the universal enveloping algebra as the limit $\hbar \to 0$ and we have that $\mathfrak D_\hbar(\mathfrak b) \cong \mathfrak D(\mathfrak b)[[\hbar]]$ as $\mathbb{C}[[\hbar]]$ modules, where $\mathfrak D(\mathfrak b)=U(\tilde \g)$.
\begin{remark}
Often in the theory of quantum groups one defines 
\begin{equation*}
   K_i^\pm = q^{H_i^\pm}, \quad q=\exp{(\hbar/2)}
\end{equation*}
for which the (non-perturbative) co-product takes the form
\begin{equation*}
   \Delta E_i=(K_i^+)^{-1/2}\otimes E_i +E_i\otimes (K_i^+)^{1/2}.
\end{equation*}
Since we are realizing the co-product in the setting of perturbation theory, it will be more convenient to use equation \eqref{eq:brackets} and \eqref{eq:generator_coprod} as our convention. 
\end{remark}
\begin{lemma}\label{lm:co-prod} The leading order correction to the co-product on a general basis element $t_a\in \mathcal B_+$ is given by
\begin{equation}
\Delta_{(1)}t_a = \frac{1}{2}\sum_{b,c=1}^{n/2}\big({f_a}^{bc}~t_{b,V}\otimes t_{c,V'}\big).
\end{equation}
Recall the definition of the structure constant:
\begin{equation}
[t_a,t_b]=\sum_{c=1}^{n}{f_{ab}}^ct_c \ , \ \ f_{abc}=\braket{[t_a,t_b],t_c}.
\end{equation}
\end{lemma}
\begin{proof}
One checks that this formula agrees with the co-product in equation \eqref{eq:generator_coprod} on the algebra generators $E_i,H^+_i$ and that it commutes with the bracket.
\end{proof}
\paragraph{The $R$-matrix}
Another part of the quasi-triangular Hopf algebra structure is an $R$-matrix element $R \in \mathfrak D_\hbar(\mathfrak b) \otimes \mathfrak D_\hbar(\mathfrak b)$ given by 
\begin{equation}
R=1 + \sum \hbar^k r^{(k)},
\end{equation}
where each $r^{(k)}$ is an element of $\mathfrak D(\mathfrak b)\otimes \mathfrak D(\mathfrak b)$. The element $r\coloneqq r^{(1)}$ is known as the classical $R$-matrix and is given~by 
\begin{equation}\label{eq:R-matrix}
r= \sum_{a=1}^{n/2} t_a \otimes t^a \,.
\end{equation}
The category of representations of $\mathfrak D_\hbar(\mathfrak b)$ is a braided monoidal category with monoidal product coming from the co-product in equation  \eqref{eq:generator_coprod} and braiding coming the $R$-matrix in equation \eqref{eq:R-matrix}. 

\section{Perturbation Theory}\label{sec:perturbation}
\subsection{The Propagator}\label{sec:propagator}
The remainder of this paper studies the expectation value of operators in the theory in the setting of perturbation theory. An essential ingredient for this is constructing a propagator, which can be thought of as the probability distribution for a gauge boson traveling between two points on the manifold. The propagator in the present setting is a Lie algebra valued two-form $ {P}\in\Omega^2\big((M\times M)\setminus\diag,\,\tilde\g\otimes\tilde\g\big)$ such that $P$ is a Green's function for the differential operator~$d$, that is 
\begin{equation}\label{eq:P-rel}
dP(x,y)=\delta^{(3)}(x,y)\,\mathscr C(\tilde\g),
\end{equation}
where $\delta^{(3)}(x,y)$ is the 3-dimensional Dirac delta distribution localized at $x=y$ and $\mathscr C(\tilde\g)\in \tilde \g\otimes \tilde \g$ is the Casimir element of $\tilde\g$.
Moreover we impose the a set of boundary conditions on the propagator coming from the boundary conditions on the gauge field in equation \eqref{eq:bc}: Write $\partial_+M$ for the upper boundary $\partial_+M=\R^2\times \{1\}$ and $\partial_-M$ for the lower boundary $\partial_-M=\R^2\times \{-1\}$. We require that
\begin{enumerate}[(i)]
    \item the restrictions $P\big |_{\partial_+M\times M}$ and $P\big |_{M\times \partial_-M}$ takes value in $\mathfrak l_+\otimes \mathfrak l_-$,
    \item the restrictions $P\big |_{\partial_-M\times M}$ and $P\big |_{M\times \partial_+M}$ takes value in $\mathfrak l_+\otimes \mathfrak l_-$.
\end{enumerate}
A two-form satisfying the equation \eqref{eq:P-rel} along with the above boundary constraints can be constructed as follows: Let $\omega=f\vol_{S^2}\in \Omega^2(S^2)$ where $\vol_{S^2}$ is the unit volume form on $S^2$ given in terms of the coordinates on $\R^3$ by 
$$
\vol_{S^2}=x\,dy\wedge dz+y\,dz\wedge dx+z\, dx\wedge dy,
$$ 
and $f\in C^\infty(S^2)$ satisfying the following properties: 
\begin{enumerate}[(i)]
    \item $f$ is only supported in a small neighbourhood of ``the north pole'' $x_{np}=(0,0,1)$
    \item $f$ is symmetric under rotations around the axis through $x_{np}=(0,0,1)$ and $x_{sp}=(0,0,-1)$
    \item $\int_{S^2}f\vol_{S^2}=1$.
\end{enumerate}
Furthermore, let $\phi:M\times M\setminus\diag \to S^2$ be the map
\begin{equation}
\phi(x,y)=\frac{y-x}{|y-x|}
\end{equation}
and let $R$ be the orientation reversing map,  $R:S^2\to S^2$, $R(x)=-x$. We now define the propagator as the pull back
\begin{equation}\label{eq:propagator}
P=\phi^*\big(\omega ~r^+-R^*\omega~r^-\big),
\end{equation}
where $r^+\in\mathfrak l_+\otimes \mathfrak l_-$ and $r^-\in \mathfrak l_-\otimes \mathfrak l_+$ are uniquely determined by the constraint in equation \eqref{eq:P-rel}. To see this, notice first that since $P$ is the pull back of a top-dimensional form on $S^2$ it holds that $dP(x,x')$ vanishes for all $x,x'$ with $x\neq x'$. Now fix $x'=0$ and consider the integral of $dP(x,0)$ when $x$ is in the unit ball around $0$. by Stokes' theorem we have
$$
\int_{x\in B}dP(x,0)=\int_{S^2}P(x,0)=\int_{S^2}\big(\omega(x)\, r^+-R^*\omega(x)\, r^-\big)=r^++r^-\,.
$$
This fixes $r^+$ and $r^-$, namely
\begin{equation}
    \begin{aligned}
    r^+=r
    \ , \ \ r^-=  T\circ r\,.
\end{aligned}
\end{equation}
where $r$ is the classical $R$-matrix given in equation \eqref{eq:R-matrix} and $T$ is the map that swaps the tensor factors. 





\subsection{Feynman Diagrams}\label{sec:Feynman}
In perturbation theory, the expectation value of an observable is computed as an expansion in the parameter $\hbar$ in terms of a set of weighted graphs (Feynman diagrams). The weight of a given graph is determined from a set of Feynman rules derived from the Chern-Simons action in equation \eqref{CS-action}. 
By a Feynman diagram in the present setting we mean the following:
\begin{definition}
    A Feynman diagram is a directed trivalent graph with leaves (external half-edges) and with the half-edges decorated by elements of $\mathcal B$ such that: A half-edge labeled by $t_a\in \mathcal B_+$ is connected by an edge to a half-edge labeled by $t^a\in \mathcal B_-$ with the edge orientation going from $\mathcal B_-$ to $\mathcal B_+$. 
\end{definition}
\paragraph{Feynman rules} The Feynman rules outlined below associates to any Feynman diagram $\Gamma$ a differential form on the space of embeddings of the vertices of $\Gamma$ into $\R^2\times I$. The weight of a Feynman diagram is computed as the integral of the associated differential form over the space of embeddings.
\begin{enumerate}[(1)]
    \item An edge going from a vertex at $p\in \R^2\times I$ to a vertex at $q\in \R^2\times I$ contributes a two-form $\hbar\,\phi^*\omega(p,q)$ coming from the propagator (see remark 2 below).
    \begin{figure}[H]
        \centering
        \includegraphics[scale=.13]{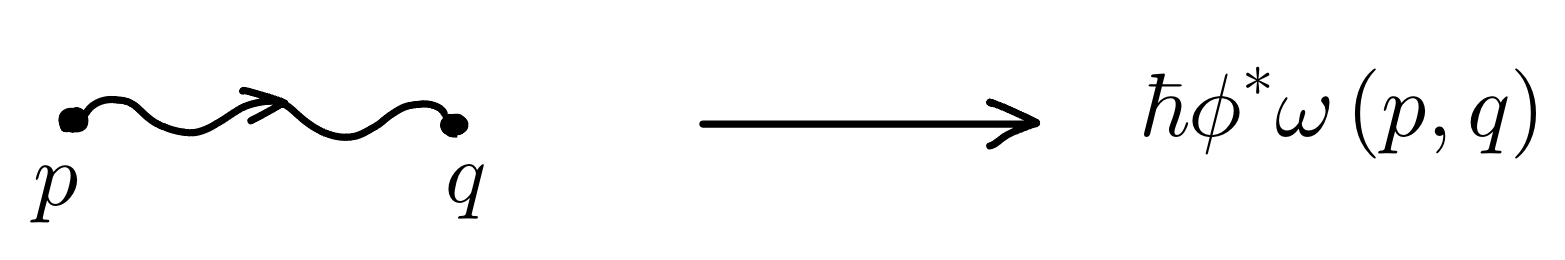}
    \end{figure}
    \item An internal vertex with incident edges labeled by basis elements $t_a,t_b,t_c\in \mathcal B$ contributes a factor structure constant $\frac{1}{\hbar}f_{abc}$.  
    \begin{figure}[H]
        \centering
        \includegraphics[scale=.11]{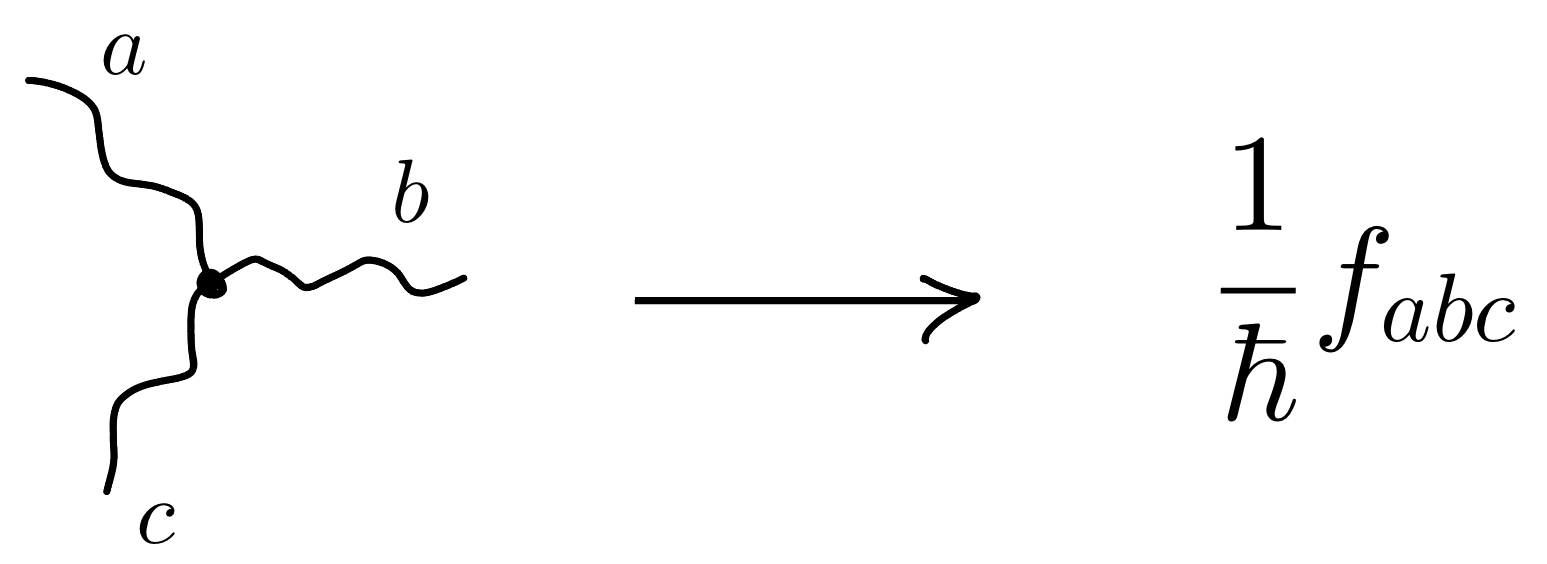}
    \end{figure} 
    Recall that the structure constant is given by 
    $f_{abc}=\braket{[t_a,t_b],t_c}$.
    \item An external half-edge labeled by $t_a\in \mathcal B$ and connected a vertex at $p\in \R^2\times I$ contributes a gauge field $A^a(p)$\,.
    \begin{remark}
    Write $P=\sum_{ab}P^{ab}t_a\otimes t_b$. We note that, in a free Chern-Simons theory (with no boundary conditions), one would consider Feynman diagrams with unoriented edges and with half-edges labeled by general elements of $\mathcal B$. To an edge with half-edges labeled by $t_a$ and $t_b$, the Feynman rules would associate the component $P^{ab}(x,y)$. However, as seen from equation \eqref{eq:propagator}, the boundary conditions in the present theory split the propagator into two parts corresponding to the two edge orientations, and we can therefore choose as a convention to define Feynman diagrams with oriented and sum over all edge orientations.
\end{remark}
\end{enumerate}

\subsection{Wilson Lines in perturbation theory}
A common set of gauge invariant observables to study in Chern-Simons theory is the so called Wilson loops. Given a closed loop $\gamma\subset M$ and a representation $V$ of $\mathfrak g$ the associated Wilson loop is defined as the trace of the holonomy of the gauge field around $\gamma$:
\begin{equation*}
\begin{aligned}
W_V(\gamma)&=\Tr_V\bigg(\mathscr P\exp \int_\gamma A\bigg) \\&\coloneqq\Tr(1_V) + \int_\gamma dx^i A_i^a(x)\,\Tr(t_{a,V}) + \int_\gamma dx^i\int^x dx'^j A^a_i(x) A^b_j(x')\,\Tr(t_{a,V}\,t_{b,V})+\dots
\end{aligned}
\end{equation*}
where $\mathscr P$ means the path ordering of the exponential and we use the notation $t_{a,V}$ to denote the basis element $t_a$ acting in the representation $V$. 
In this paper, we consider instead a set of operators called Wilson lines coming from omitting the trace and replacing the closed loop $\gamma$ with an open line $L$ extending to infinity along $\R^2$. In the setting of perturbation theory, we think of a Wilson line $L(V)$ simply as a pair $(L,V)$, and we allow a gauge field $A^a$ to couple to $L(V)$ by inserting a basis element $t_{a,V}$ at the corresponding point on $L$. In other words, we expand the definition of Feynman diagrams to include graphs with univalent vertices along $L$, with the additional Feynman rule that a vertex on $L$ with incident half-edge labeled by $t_a$ contributes an element $t_{a,V}$. 
    \begin{figure}[H]
    \centering
    \includegraphics[scale=.11]{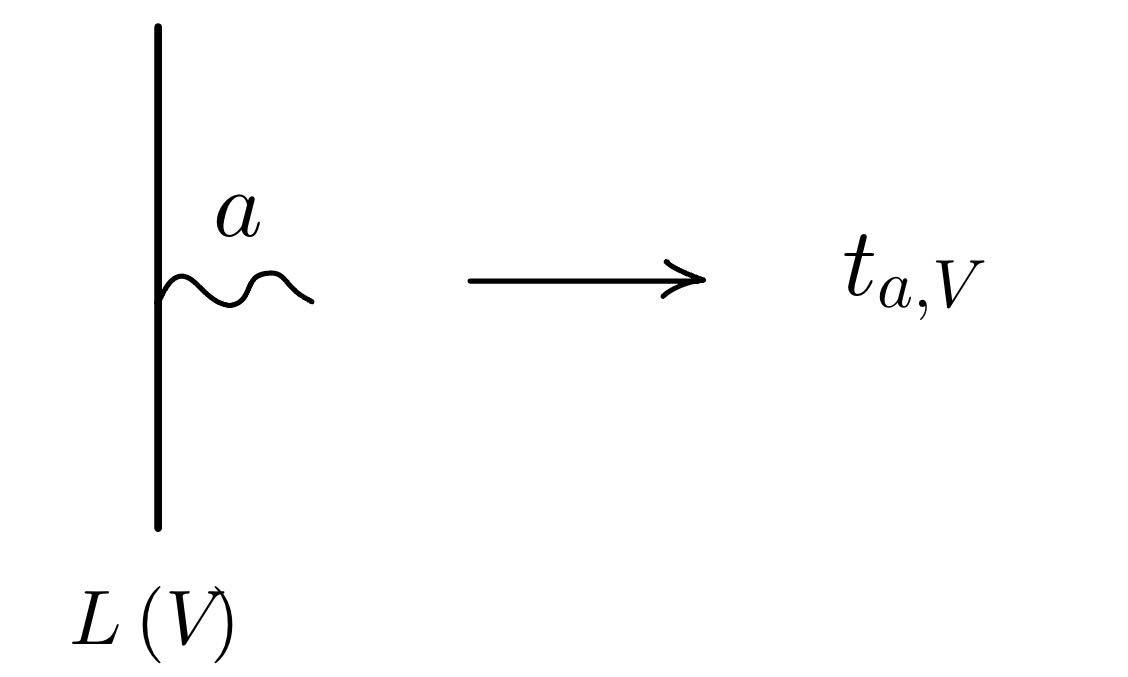}
    \caption{: Feynman rule for the coupling of a gauge field $A^a$ to the Wilson line $L(V)$.}
\end{figure}

\section{Quantum Groups and Wilson lines}\label{sec:Main}
\subsection{Merging parallel Wilson lines}\label{sec:stating}
The study of the remainder of the paper will be the product of two parallel Wilson line operators in the limit when the lines come close together. We fix a set of coordinates $(x,y,z)$ on $\R^2\times I$ with $(x,y)$ coordinates in $\R^2$ and $z$ the coordinate along $I$. Let $L(V)$ be a Wilson line supported at $x=z=0$ and $L_\varepsilon(V')$ a Wilson line supported at $x=\varepsilon$, $z=0$. We write $L(V)L_\varepsilon(V')$ to mean the disjoint union of the lines $L$ and $L_\varepsilon$ such that a gauge field $A^a$ couples to the line $L$ by inserting an element $t_{a,V}\otimes 1_{V'}$ and to the line $L'$ by inserting an element $1_V\otimes t_{a,V'}$. In general, the coupling of an external gauge field to the two Wilson lines is given by a perturbative expansion in $\hbar$ using the Feynman rules in section \ref{sec:Feynman}:
\begin{equation}\label{eq:coupling}
    \mathcal A_a\big(L(V)L_\varepsilon(V')\big) =\sum_{k=0}^\infty\hbar^k\mathcal A_a^{(k)}\big(L(V)L_\varepsilon(V')\big) \in \End(V\otimes V')
\end{equation}
where each element $\mathcal A_a^{(k)}\big(L(V)L_\varepsilon(V')\big) \in \End(V\otimes V')$ is computed as the weighted sum over Feynman diagrams with a single external half-edge (leaf) labeled by $t_a$ and with the number of internal edges minus the number of internal vertices equal to $k$. In the limit $\varepsilon\to 0$ one would expect equation \eqref{eq:coupling} to reproduce the expression for an external gauge field coupling to a single Wilson line at $L$. It is however not immediately clear what representation should be associated to the merged Wilson line.
\vspace{.3cm}
\begin{figure}[H]
    \centering
    \includegraphics[scale=.14]{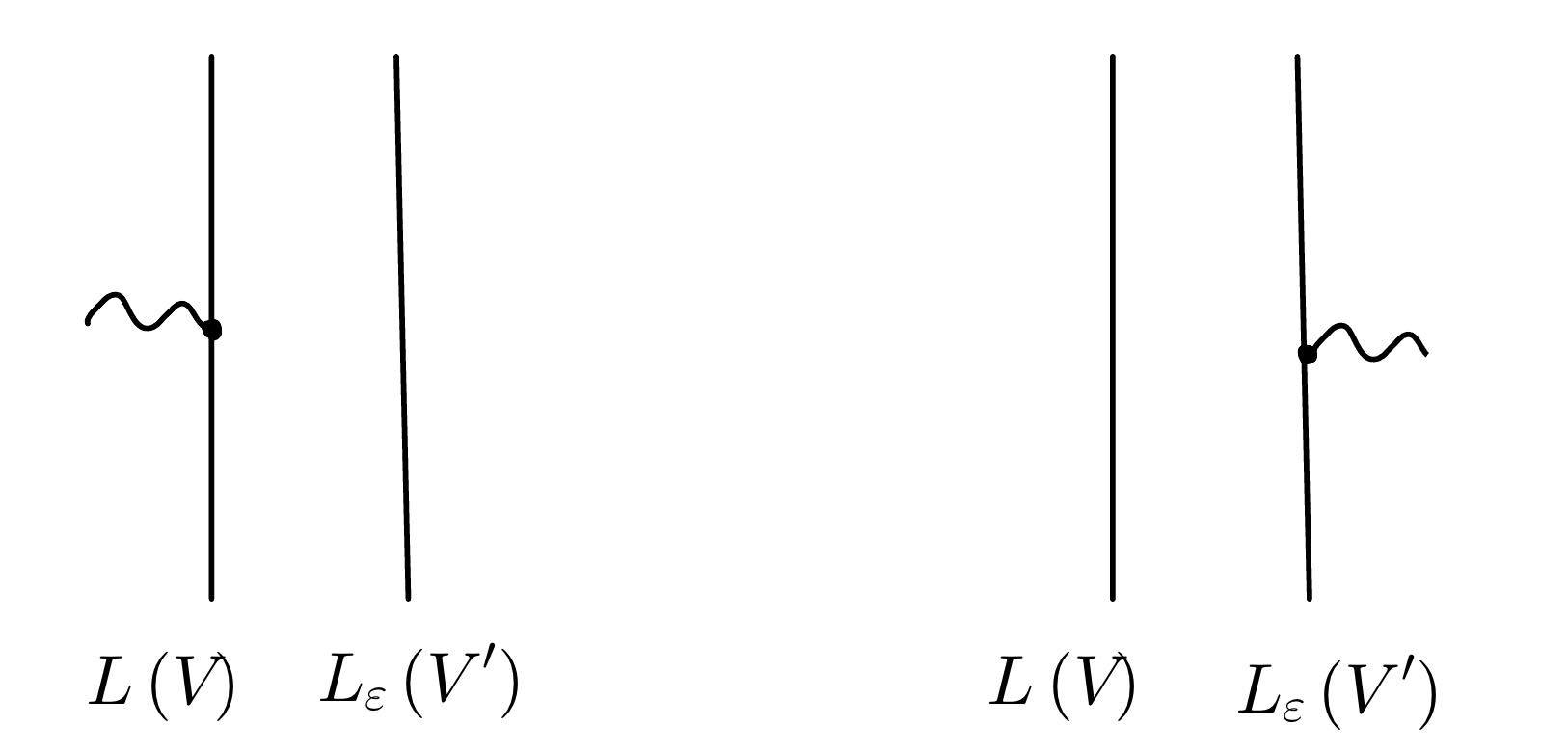}
    \caption{:~ The classical level Feynman diagrams for an external gauge field coupling to the two Wilson lines.}
    \label{fig:classical}
\end{figure}
At the classical level the gauge field simply couples to each line individually as shown in figure~\ref{fig:classical}, and the corresponding Feynman amplitude is given by
$$
\mathcal A^{(0)}_a\big(L(V) L_\varepsilon(V')\big)=\int_{q\in L}A^{a}(q)\,t_{a,V}\otimes 1_{V'}+\int_{q'\in L_\varepsilon}A^a(q')\,1_{V}\otimes t_{a,V'}.
$$
Taking the limit $\varepsilon\to 0$ on the right-hand side in the above we get
$$
\int_{q\in L}A^a(x)\,\big(t_{a,V}\otimes 1_{V'}+1_V\otimes t_{a,V'}\big)\,,
$$
which is the expression for a gauge field coupling to a single Wilson line at $L$ in the tensor product representation $V\otimes V'$. Hence, at the classical level we have
\begin{equation}\label{eq:classical}
\begin{aligned}
\lim_{\varepsilon\to 0}\mathcal A^{(0)}_{a}\big(L(V) L_\varepsilon(V')\big)=\mathcal A^{(0)}_a\big(L(V\otimes V')\big).
\end{aligned}
\end{equation}
The object of the remainder of this paper is to carry out the computation of the leading order contribution $\mathcal A_{a}^{(1)}\big(L(V)L_\varepsilon(V')\big)$ in the limit $\varepsilon\to 0$. As we shall see, this gives a correction to the tensor product $V\otimes V'$ in equation \eqref{eq:classical} which agrees with the leading order quantum deformation of the tensor product in $\mathfrak D_\hbar(\mathfrak b)$ given in equation \eqref{eq:generator_coprod}. This is expressed in the following theorem:
\begin{theorem}\label{main statement} It holds that 
$$
\lim_{\varepsilon\to 0}\mathcal A_a^{(1)}\big(L(V) L_\varepsilon(V')\big)=\mathcal A_a^{(1)}\big(L(V\otimes_\hbar V')\big),
$$
where $V\otimes_\hbar V'$ is the tensor product in $\Rep \mathfrak D_\hbar(\mathfrak b)$ defined via the co-product in equation~\eqref{eq:generator_coprod}.
\end{theorem}
Notice that lemma \ref{lm:co-prod} in section \ref{sec:quantization} defines the relevant co-product on a general basis element $t_a\in \mathcal B_+$ and it follows that:
\begin{equation}\label{eq:thm1}
\mathcal A_a^{(1)}\big(L(V\otimes_\hbar V')\big)=\frac{1}{2}\sum_{b,c=1}^{\dim\tilde \g/2}\big({f_a}^{bc}~t_{b,V}\otimes t_{c,V'}\big) \int_{q\in L}A^a(q).
\end{equation}

We conjecture theorem \ref{main statement} to hold at all orders in perturbation theory. However, explicitly computing the contributing Feynman integrals at higher orders appears to be too difficult a task, and a proof would therefore require different techniques. 
\subsection{The Configuration Space of Vertices}\label{sec:configuration}
The differential form associated to a Feynman diagram $\Gamma$ is defined on the configuration space of vertices of $\Gamma$. We here give a definition of the relevant configuration space in the presence of Wilson lines $L,L_\varepsilon$, and we refer the reader to \cite{Bott1994OnTS} for a more general definition. As we are interested in studying the limit when $\varepsilon$ goes to $0$ it will be convenient to think of $\varepsilon$ as a parameter in the configuration space. 
\begin{definition} \label{def:conf} For $n_1, n_2, m\in \mathbb Z_{\geq 0}$ define $\Conf_{n_1,n_2,m}$ to be the space of points 
$$\{\varepsilon,q_1,\dots q_{n_1},q'_1,\dots, q'_{n_2},p_{1},\dots, p_m\},$$  
where $\varepsilon\in [0,\infty)$ and 
\begin{align*} 
&q_1,\dots q_{n_1}\in L \ \text{ with } \ q_i\neq q_j\,, \\ &q'_1,\dots, q'_{n_2}\in L_\varepsilon \ \ \text{ with } \ q'_i\neq q'_j\,, \\ 
&p_1,\dots, p_m\in (\R^2\times I)\setminus \{q_1,\dots,q_{n_1},q'_1,\dots, q'_{n_2}\}
\ \text{ with } \ p_i\neq p_j.
\end{align*}
Furthermore, consider the projection onto the first factor
$$
\Conf_{n_1,n_2,m}\to (0,\infty).
$$
We denote by $\Conf_{n_1,n_2,m}^\varepsilon$ the fiber of this projection over a fixed $\varepsilon\in(0,\infty)$.
\end{definition}

\subsection{Contributing Diagrams at Leading Order}\label{sec:leading_order}
\begin{figure}[H]
    \centering
    \includegraphics[scale=.18]{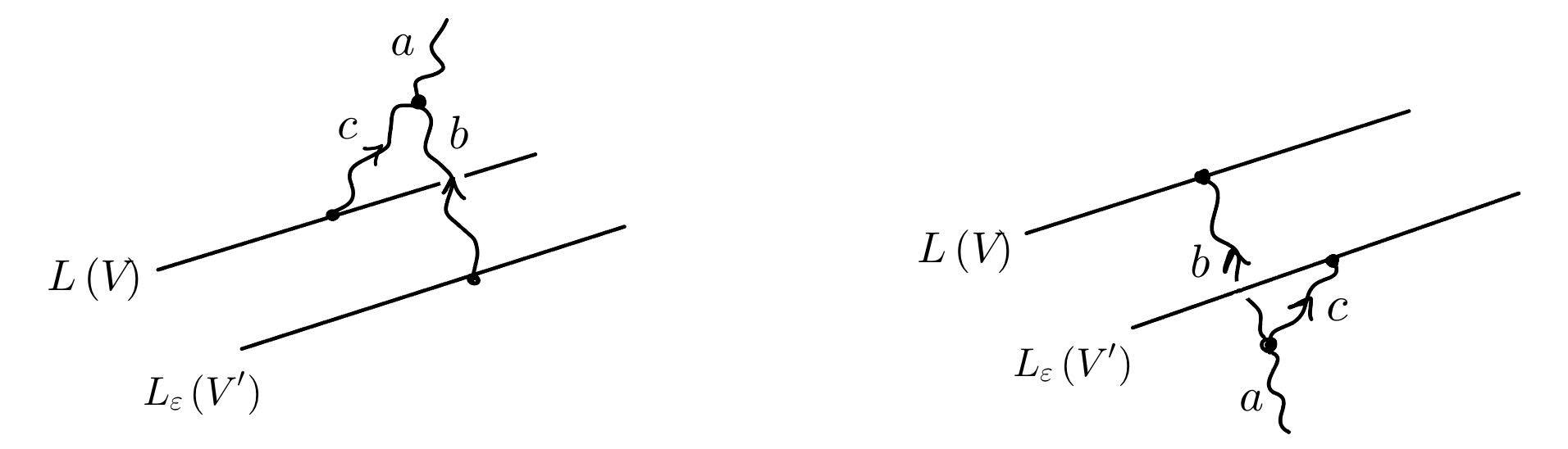}
    \caption{:~ Contributing diagrams.}
    \label{fig:leading_order}
\end{figure}
\begin{figure}[H]
    \centering
    \includegraphics[scale=0.175]{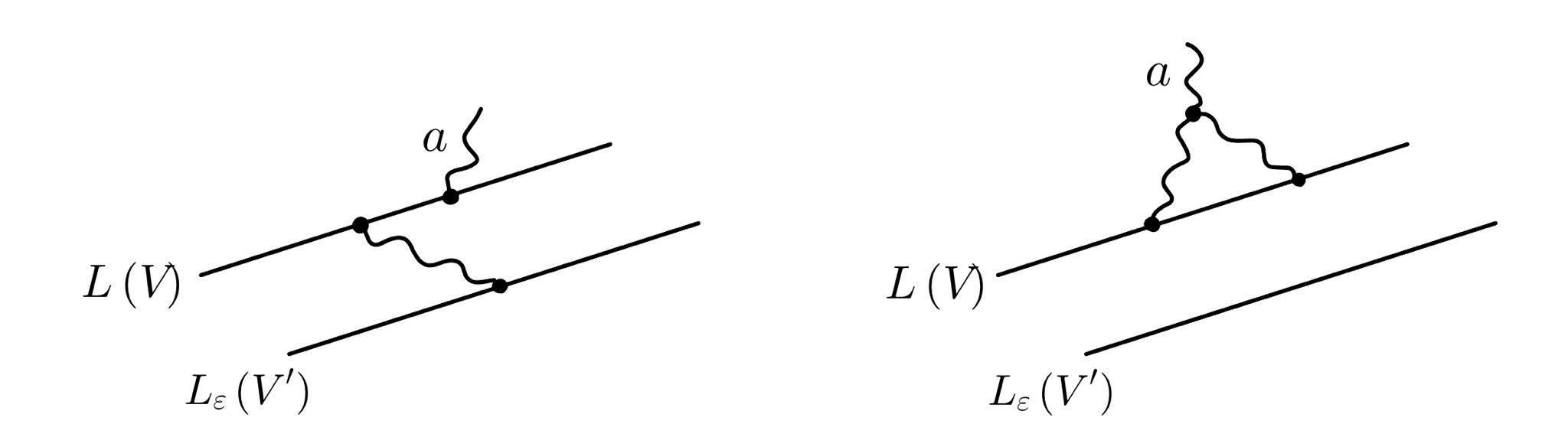}
    \caption{:~ Vanishing diagrams.}
    \label{fig:vanishing}
\end{figure}
\begin{lemma}The only Feynman diagrams contributing to $\mathcal A^{(1)}_a\big(L(V)L_\varepsilon(V')\big)$ are the ones shown in figures~\ref{fig:leading_order}.
\end{lemma}
\begin{proof}
Recall that the diagrams contributing to $\mathcal A_a^{(1)}\big(L(V)L_\varepsilon(V)\big)$ has the number of internal edges minus the number of internal vertices equal to one. This gives precisely the diagrams shown in figure \ref{fig:leading_order} and \ref{fig:vanishing}. Consider first the diagram on the left-hand side of figure \ref{fig:vanishing}. Since the lines are in the same plane parallel to the boundary, the form vanishes due to the propagator only being non-zero in a small neighbourhood of the north pole. Consider now diagram on the right-hand side of figure \ref{fig:vanishing}. The associated configuration space is $\Conf_{2,0,1}$. Let $G<\Homeo(\R^3)$ be the subgroup of scalings and translations along $L$ and consider the quotient map:
\begin{equation}\label{eq:quotient}
\Conf_{2,0,1}\to  \Conf_{2,0,1}/G
\end{equation}
The subgroup $G$ is two-dimensional and hence the space $\Conf_{2,0,1}/G$ has dimension $5-2=3$. On the other hand, let $P_1\wedge P_2\in\Omega^4(\Conf_{2,0,1})$ be the product of propagators associated to the internal edges, that is, 
$$P_1\wedge P_2\,(q_1,q_2,p)\coloneqq\phi^*\omega(q_1,p)\wedge \phi^*\omega(q_2,p).$$ By definition the propagator is invariant under scalings and translations along the $L$ and hence the form $P_1\wedge P_2$ factors through the quotient map in equation \eqref{eq:quotient}. By dimensional counting, this implies that $P_1\wedge P_2$ vanishes.
\end{proof}
Consider therefore the diagrams in figure \ref{fig:leading_order}. We can assume that $t_a \in \mathcal B_+$ since the computation for $t_a\in \mathcal B_-$ is entirely analogous. In this case, the only contribution to the expectation value comes from the diagram on the right-hand side of figure~\ref{fig:leading_order}. In fact, the internal vertex of diagram on the left-hand side of figure \ref{fig:leading_order} has all three incident half edges labeled by elements $t_a,t_b,t_c\in \mathcal B_+$. By the Feynman rules in section \ref{sec:Feynman} this vertex is assigned a structure constant $f_{abc}=\braket{[t_a,t_b],t_c}$ which is zero since the Killing form vanishes on $\mathfrak l_+$. 
The Feynman amplitude coming from the diagram on the right-hand side of figure \ref{fig:leading_order} takes the form
\begin{equation}\label{eq:leading_order}
\begin{aligned}
\mathcal A_a^{(1)}\big(L(V) L_\varepsilon(V')\big)&=\sum_{b,c=1}^{\dim\tilde \g/2}\big({f_a}^{bc}~t_{b,V}\otimes t_{c,V'}\big)~\mathcal I_\varepsilon\,,
\end{aligned}
\end{equation}
where
\begin{equation}\label{eq:I_e}
\mathcal I_\varepsilon\coloneqq\int_{\Conf_{1,1,1}^\varepsilon}A^a(p)\wedge \phi^*\omega(p,q)\wedge \phi^*\omega(p,q').
\end{equation}
\subsection{A Configuration Space Compactification} \label{sec:compactification}
\begin{figure}[H]
    \centering
    \includegraphics[scale=0.15]{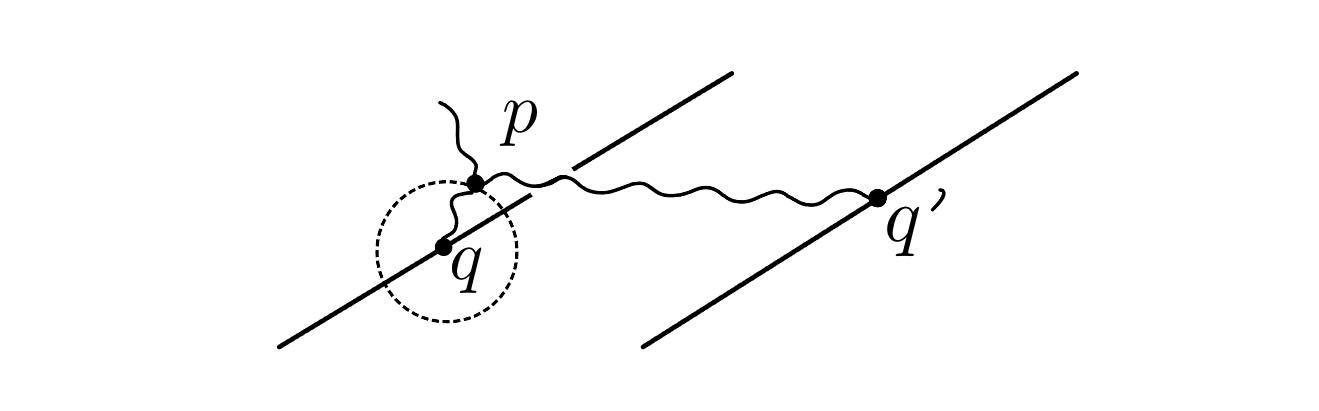}
    \caption{: Neighbourhood of $\Conf_{1,1,1}$ where $p$ is close to $q$ and far from $q'$.}
    \label{fig:vanishing_strata}
\end{figure}

Since the propagator is only defined away from the diagonal, it is not clear what will happen to the integral $\mathcal I_\varepsilon$ in equation \eqref{eq:I_e} when $p\to q, q'$. In order compute $\lim_{\varepsilon\to 0}\mathcal I_\varepsilon$, we must therefore define a partial compactification $\overline{\Conf}_{1,1,1}$ of ${\Conf}_{1,1,1}$ in the direction $\varepsilon\to 0$ such that the integrand extends smoothly to the corresponding boundary $\partial_\varepsilon\,\overline{\Conf}_{1,1,1}$. Then we have 
\begin{equation}\label{eq:lim}
\lim_{\varepsilon\to 0} \mathcal I_\varepsilon= \int_{\partial_\varepsilon\overline \Conf_{1,1,1}}A^a(p)\wedge \phi^*\omega(p,q)\wedge \phi^*\omega(p,q')\,.
\end{equation}
To this aim, we use the so called Fulton-MacPherson configuration space compactification. This compactification was originally due to Fulton and MacPherson \cite{fultonMacpherson} and applied to Chern-Simons perturbation theory by Axelrod and Singer \cite{axelrod1994chern} and (in the presence of Wilson loops) Bott and Taubes~\cite{Bott1994OnTS}. In this compactification $\partial_\varepsilon\,\overline{\Conf}_{1,1,1}$ can be divided into the disjoint union of strata
$$
\partial_\varepsilon\, \overline{\Conf}_{1,1,1}=\bigcup_{i=1}^3\partial_i\,\overline{\Conf}_{1,1,1}
$$
corresponding to the following cases:
\begin{enumerate}[(a)]
    \item The internal vertex $p$ remains far from the lines compared to $\varepsilon$ as $\varepsilon\to 0$.
    \item The internal vertex $p$ moves close to the lines and at least one vertex $q$ or $q'$ remains far from $\varepsilon\to 0$.
    \item All three vertices move close to each other as $\varepsilon\to 0$.
\end{enumerate}
\begin{lemma}\label{lm:1}
We get no contribution to equation \eqref{eq:lim} coming from the boundary stratum $\partial_1\,\overline{\Conf}_{1,1,1}$ corresponding to case (a) in the above. 
\end{lemma}
\begin{proof}
When $p$ is far from the lines the integrand in equation \eqref{eq:I_e} extends smoothly to the boundary coming from allowing $\varepsilon\to 0$, and the corresponding boundary stratum takes the form
$$
\partial_1\overline{\Conf}_{1,1,1}=\big\{(q,q',p)\in L\times L\times (\R^2\times I\setminus L)~\big|~p\neq q,q'\big\}\,.
$$
The contribution to equation $\eqref{eq:lim}$ is given by
\begin{equation}
\int_{p\in (\R^2\times I)\setminus L}A(p)\wedge \bigg(\int_{q\in L}\phi^*\omega(q,p)\bigg)\wedge\bigg(\int_{q'\in L}\phi^*\omega(q',p)\bigg)\,,
\end{equation}
which is zero since the last two factors are identical one forms. 
\end{proof}
\begin{lemma}\label{lm:2} We get no contribution to equation \eqref{eq:lim} coming from the boundary stratum $\partial_2\,\overline{\Conf}_{1,1,1}$ corresponding to case (b) in the above. 
\end{lemma}
\begin{proof} This follows from the property that $\omega$ is only non-zero in a small neighbourhood of the north pole. In fact, assume that $p$ is approaching some point $q\in L$ and that $q'$ remains far from $p$ as $\varepsilon\to 0$. Because the Wilson lines are in the same plane parallel to the boundary it holds that, given any $\eta>0$ there is a $\delta>0$ such that, if we define $U\subset \Conf_{1,1,1}$ to be the neighbourhood where $|p-q'|>\eta$ and $|p-q|<\delta$ then $\phi^*\omega(p,q')=0$ for all $(p,q,q')\in U$. The situation is illustrated in figure \ref{fig:vanishing_strata}
\end{proof}
By lemma \ref{lm:1} and \ref{lm:2}, the only contribution to equation \eqref{eq:lim} comes from the boundary stratum $\partial_3\Conf_{1,1,1}$ corresponding to all three vertices coming together as $\varepsilon\to 0$. To define the corresponding boundary stratum we need the following definition:
\begin{figure}[H]
    \centering
    \includegraphics[scale=0.08]{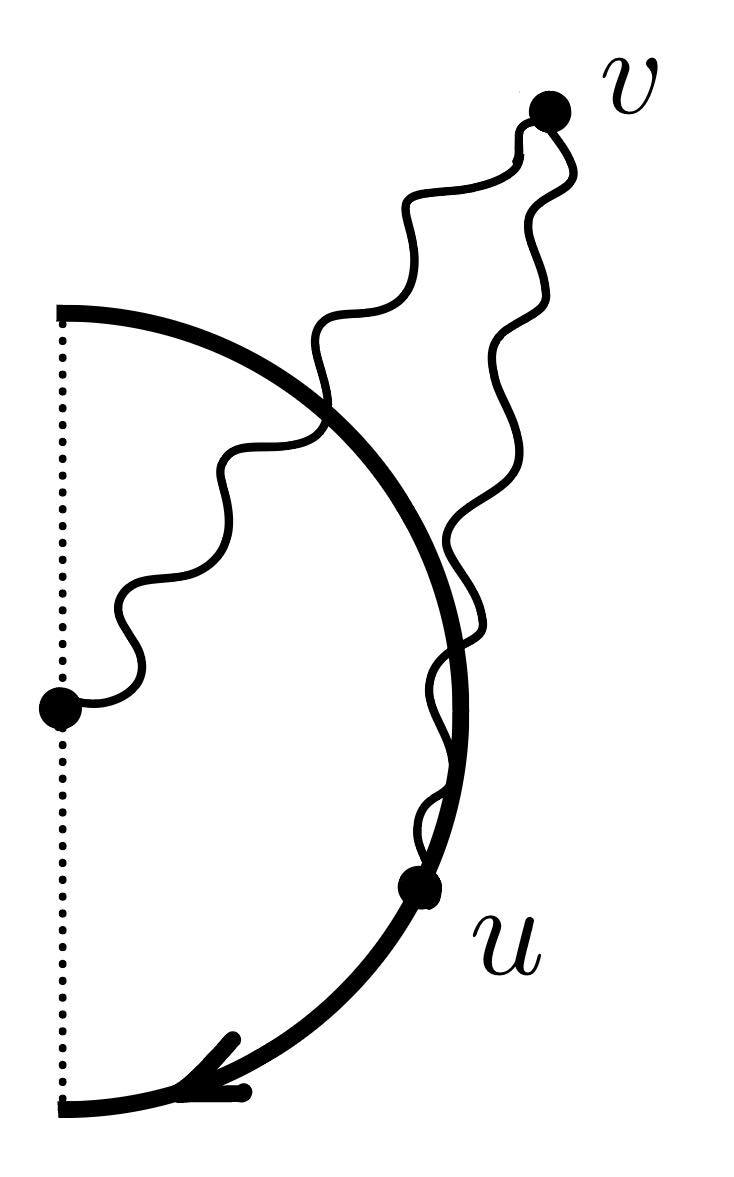}
    \caption{: The space $\mathcal H$.}
    \label{fig:half_circ}
\end{figure}
\begin{definition}\label{def:conftilde} Let $S_r$ be the ``right'' half of the unit circle with center $(0,0)$, i.e.
$$
S_r = \big \{(x,y,0)\in \R^2\times I ~ \big|~x^2+y^2=1 \, , \ x >0 \big \} 
$$ 
We define
$$
\mathcal H=\big\{(u,v)\in S_r\times \R^3\setminus \{0\}~\big|~v\neq u\big\}.
$$
\end{definition}
The space $\mathcal H$ is illustrated in figure \ref{fig:half_circ}.
\begin{lemma} \label{lm:3} We can define a partial compactification of $\Conf_{1,1,1}$ in the direction where all three vertices come together, such that the compactified space is a manifold with boundary and corresponding boundary stratum $\partial_3\,\overline{\Conf}_{1,1,1}$ is given by
    $$\partial_3\, \overline{\Conf}_{1,1,1}=L\times \mathcal H\,.$$
\end{lemma}
\begin{proof}
    For some small $\eta>0$, define $U\subset \Conf_{1,1,1}$ by
    $$
    U=\big \{(\varepsilon,q,q',p)\in \Conf_{1,1,1}~\big |~ |p-q|<\eta \text{ and }  |q'-q|<\eta\big \}.
    $$
    Furthermore, define $V\subset(0,\eta)\times L\times \mathcal H$ by
    \begin{equation}\label{eq:V}
     V= \big\{(t,q_0,(u,v))\in (0,\eta)\times L\times \mathcal H~\big |~|v|t<\eta\big\},
    \end{equation}
    There exists a diffeomorphism $\varphi:V\to U$ defined by 
    $(t,q_0,(v,u))\mapsto (\varepsilon,q,q',p)$, where
\begin{equation}\label{eq:coord_change}
\begin{aligned}
    \varepsilon=u_y \ , \ \ q=q_0 \ , \ \ q'=q_0+t u\ , \ \ p = q_0+t v\,,
\end{aligned}
\end{equation}
with $u_y$ denoting the $y$-coordinate of $u$. This implies that
$$
\overline{\Conf}_{1,1,1}\coloneqq\Conf_{1,1,1}\cup_V \overline V,
$$
where 
$$
\overline V= \big\{(t,q_0,(u,v))\in [0,\eta]\times L\times \mathcal H~\big |~|v|t<\eta\big\},
$$
is a manifold with boundary. Letting all three vertices come together in $\Conf_{1,1,1}$ corresponds to letting $t\to 0$ in $\overline{V}$ and the lemma follows.
\end{proof}
\subsection{Proof of Theorem 1}\label{sec:computing}
We are now equipped to prove theorem~\ref{main statement}. Notice first that with the change of coordinates given in equation \eqref{eq:coord_change}, we have
$$
A(p)=A(q+tv)\ , \ \ \phi^*\omega(p,q)=
\phi^*\omega(0,v)\ , \ \ \phi^*\omega(p,q')=
\phi^*\omega(v,u).
$$
All of the above forms extends continuously to the boundary $\partial_3\,\overline{\Conf}_{1,1,1}$ corresponding to the limit $t\to 0$. Hence, by equation \eqref{eq:lim} and lemma \ref{lm:1}, \ref{lm:2} and \ref{lm:3} in the previous subsection, it holds that
\begin{equation}\label{eq:half_circ}
\begin{aligned}
    \lim_{\varepsilon\to 0}\mathcal I(\varepsilon)
    =\int_{q\in L}A^a(q)\int_{(u,v)\in\mathcal H}\phi^*\omega(v,0)\wedge \phi^*\omega(v,u).
\end{aligned}
\end{equation}
By equation \eqref{eq:thm1} and \eqref{eq:leading_order}, proving theorem \ref{main statement} now amounts to showing that the second integral in equation \eqref{eq:half_circ} contributes a factor of $1/2$. This is the goal of the present subsection.
\begin{figure}[H]
    \centering
    \includegraphics[scale=0.12]{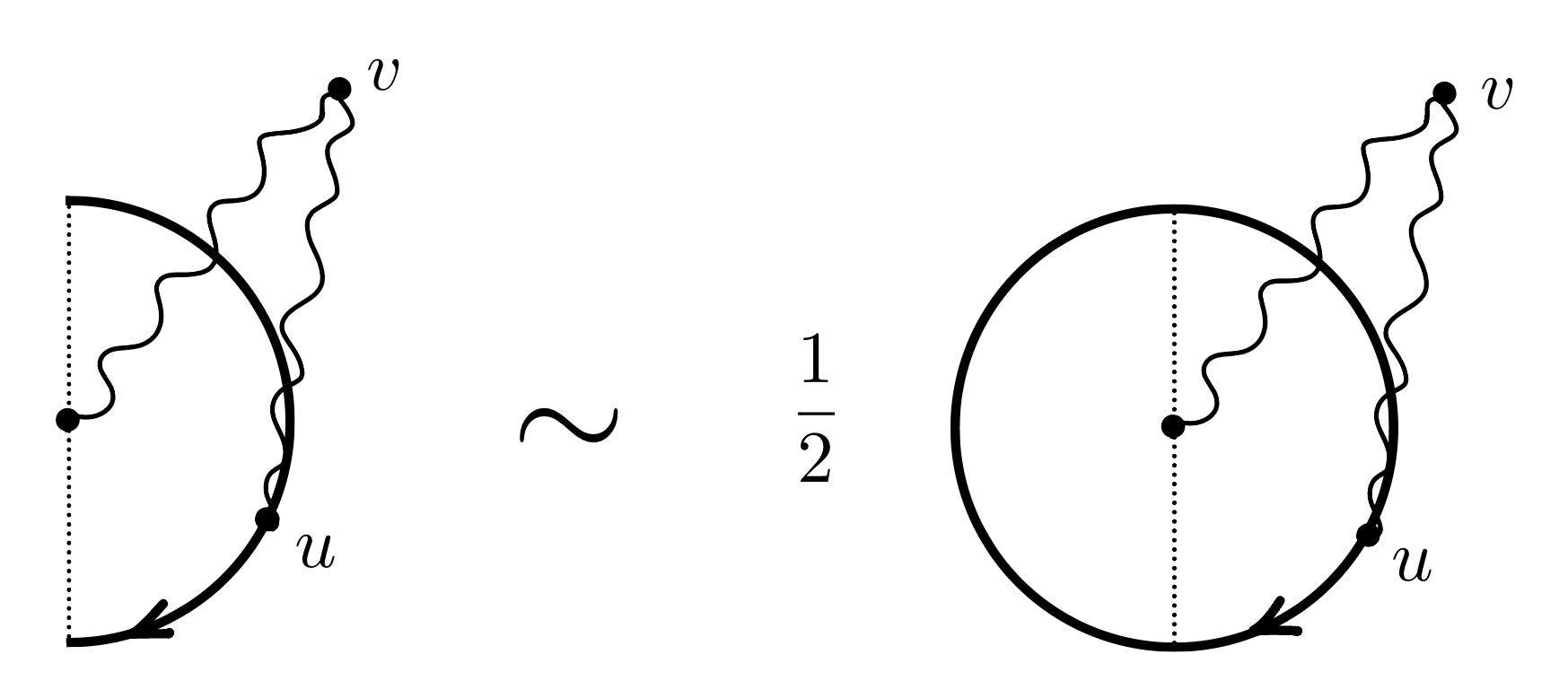}
    \caption{: The space $\C$.}
    \label{fig:full_circle}
\end{figure}
\begin{proof}[Proof of theorem 1] Let $\C\supset \mathcal H$ be the spaces obtained from allowing $u$ to be in the full circle $S\subset \R^2\times \{0\}$. That is, we define
$$
\C = \big \{(u,v)\in S\times \R^3\}\,.
$$
Due to the rotation symmetry of $\omega$ (see section \ref{sec:propagator}), it holds that 
\begin{equation}\label{eq:full_circ}
\int_{(u,v)\in\mathcal H} \phi^*\omega(v,0)\wedge \phi^*\omega(v,u)=\frac{1}{2}\int_{(u,v)\in\C} \phi^*\omega(v,0)\wedge \phi^*\omega(v,u).
\end{equation}
In fact, we are going to modify the space of integration even further: Recall from section \ref{sec:propagator} that $\omega$ is only supported in a small neighbourhood of the north pole. Hence, the only contribution to the integral in the right-hand side of equation \eqref{eq:full_circ} comes from when $v\in \R^3_-\coloneqq\R^2\times (-\infty,0)$. Defining $\C_-\subset \C$ by
$$
\C_- =\{(u,v)\in S\times \R^3_-\}\,,
$$
we can therefore replace $\C$ with $\C_-$ in equation \eqref{eq:full_circ}:
\begin{equation}\label{eq:full_circ_-}
\int_{(u,v)\in\mathcal H} \phi^*\omega(v,0)\wedge \phi^*\omega(v,u)=\frac{1}{2}\int_{(u,v)\in\C_-} \phi^*\omega(v,0)\wedge \phi^*\omega(v,u).
\end{equation}
The integral on the right-hand side of equation \eqref{eq:full_circ_-} can be computed using purely geometric arguments. Let $S_+^2$ be the upper half of the unit sphere and consider the map
$\Phi:\C_-\to S^2_+\times S^2_+\setminus \diag$, given by 
$$
\Phi(u,v)=(\phi(v,0),\phi(u,v))=\left(-\frac{v}{|v|},\frac{u-v}{|u-v|}\right).
$$
\begin{lemma}\label{lm:Phi}
    The map $\Phi$ is a diffeomorphism. 
\end{lemma}
\begin{proof}
An inverse map $\Phi^{-1}$ is constructed as follows: Let $(a,b)\in S^2_+\times S^2_+\setminus \diag$. For any $u\in S$ write $r^u_a$ for the ray going out from $u$ and pointing along the vector $-a$ and write $r_b$ for the ray going out from $0$ and pointing along the vector $-b$. Because $a\neq b$, as we move $u$ around the circle we encounter exactly one point $u_{ab}$ for which the rays $r^u_a$ and $r_{b}$ intersect. Denoting the corresponding point of intersection by $v_{ab}$ we obtain an inverse map by defining $\Phi^{-1}(a,b)=(u_{ab},v_{ab})$.
\end{proof}
From lemma \ref{lm:Phi} and the property that $\omega$ integrates to one on $S^2$ it now follows that
\begin{equation*}
\begin{aligned}
\int_{\C_-}\phi^*\omega(v,0)\wedge \phi^*\omega(v,u)=\int\displaylimits_{S^2_+\times S^2_+}\omega(a)\wedge\omega(b)=1.
\end{aligned}
\end{equation*}
Notice that we can include the diagonal $\diag\subset S^2_+\times S^2_+$ in the integral because $\omega$ extends continuously to the diagonal which is a subspace of co-dimension one. Inserting this back into equation \eqref{eq:half_circ} we get 
$$
\lim_{\varepsilon\to 0}\mathcal I_\varepsilon=\frac{1}{2}\int_{q\in L}A(q)\,.
$$
By equation \eqref{eq:thm1} and \eqref{eq:leading_order} this completes the proof of theorem \ref{main statement}.
\end{proof}


\section{Outlook to moduli spaces of local systems}\label{sec:moduli}

To a Lie group $G$ and a surface $S$ with punctures, boundaries, and marked points on the boundaries, Goncharov and Shen \cite{goncharov2022quantum} construct a moduli space $\loc{S}{G}$ which parameterizes $G$ local systems on $S$ along with some extra data at the punctures, boundaries, and marked points of $S$. These spaces are closely related to the $\mathcal{X}$ moduli spaces of ``framed''  local systems on $S$ originally constructed by Fock and Goncharov \cite{FockGonch:1}, with a slight modification to allow for cutting and gluing of surfaces. 
In particular, one associates to each marked point on the boundary the conjugacy class of the Borel subgroup $B\subset G$. 
\begin{figure}[H]
    \centering
    \includegraphics[scale=.17]{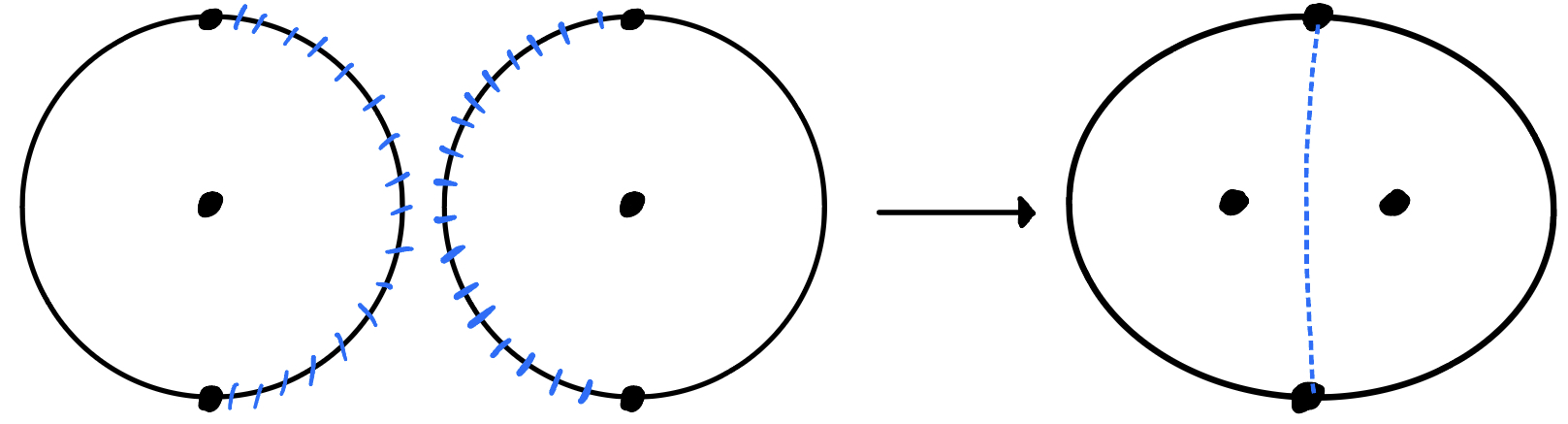}
    \caption{: The gluing map on surfaces.}
    \label{fig:glueing}
\end{figure}

\begin{figure}[b]
    \centering
    \includegraphics[scale=.17]{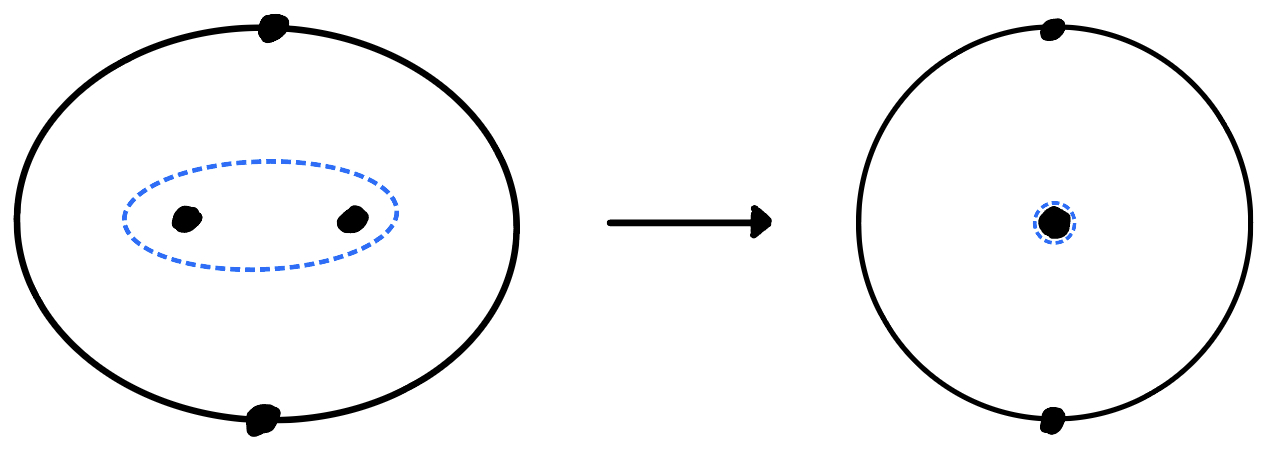}
    \caption{: The cutting map.}
    \label{fig:cutting}
\end{figure}
There is a quantization of the ring of regular functions on $\loc{S}{G}$ which is denoted by $\mathcal{O}_\hbar(\loc{S}{G})$. Goncharov and Shen construct a natural gluing operation on these quantized spaces coming from gluing surfaces along the boundary segments between two marked points; when $S$ is obtained by gluing $S'$ and $S''$ along boundary segments between marked points one obtains a map 
$$ \mathcal{O}_\hbar(\loc{S}{G}) \xrightarrow{Glue} \mathcal{O}_\hbar(\loc{S'}{G})\otimes \mathcal{O}_\hbar(\loc{S''}{G}). $$
To see how this construction relates that of the present paper, consider a disk with one puncture and two marked points on its boundary. There is a map 
$$
\kappa: \mathfrak D_\hbar(\mathfrak b) \to \mathcal{O}_\hbar(\Loc)
$$ 
which is given explicitly on generators, see \cite{SchraderShapiro} for a very nice exposition on this in the $\mathfrak{sl}_n$ case and see \cite{shen2022cluster} for the general ADE case. The coproduct is given in $\mathcal{O}_\hbar(\Loc)$ as follows: Take two copies of the punctured disk and glue them along boundary segments between marked points to obtain a twice punctured disk with two marked points on its boundary (see figure \ref{fig:glueing}). We denote the twice punctured disk by $T$. By bringing the two punctures close together and cutting out a small circle around the two punctures one obtains a new once punctured disk (see 
figure \ref{fig:cutting}). This construction gives a map 
\begin{equation}
      \mathcal{O}_\hbar(\Loc) \xrightarrow{Cut} \mathcal{O}_\hbar(\loc{T}{G}) \xrightarrow{Glue} \mathcal{O}_\hbar(\Loc) \otimes \mathcal{O}_\hbar(\Loc).
\end{equation}
which agrees with the coproduct in $\mathfrak D_\hbar(\mathfrak b)$. Similarly, the braiding on $\mathfrak D_\hbar(\mathfrak b)$ is given on $\mathcal{O}_\hbar(\Loc)$ as the map twisting the two punctures around each other:
\begin{equation}
    \mathcal{O}_\hbar(\loc{T}{G})  \xrightarrow{Braid} \mathcal{O}_\hbar(\loc{T}{G})
\end{equation}
In other words there are commutative diagrams:
\begin{align*}
       & \hspace{7em} \begin{tikzcd}[ampersand replacement = \&]
        \mathfrak D_\hbar(\mathfrak b)\ar[r, "\Delta"]\ar[d,"\kappa"] \&   \mathfrak D_\hbar(\mathfrak b) \otimes \mathfrak D_\hbar(\mathfrak b) \ar[d,"\kappa \otimes \kappa"]\\
        \mathcal{O}_\hbar(\Loc) \ar[r,"Glue"] \&  \mathcal{O}_\hbar(\Loc) \otimes \mathcal{O}_\hbar(\Loc)
    \end{tikzcd} \\ \\
    &\begin{tikzcd}[ampersand replacement = \&]  
        \mathfrak D_\hbar(\mathfrak b)\ar[r, "\Delta"]\ar[d,"\kappa"] \&    
        \mathfrak D_\hbar(\mathfrak b) \otimes \mathfrak D_\hbar(\mathfrak b)\ar[rr, "R"] \& \&   \mathfrak D_\hbar(\mathfrak b) \otimes \mathfrak D_\hbar(\mathfrak b) \ar[d,"\kappa \otimes \kappa"]\\
        \mathcal{O}_\hbar(\Loc) \ar[r,"Cut"] \&  \mathcal{O}_\hbar(\loc{T}{G})\ar[r,"Braid"] \& \mathcal{O}_\hbar(\loc{T}{G}) \ar[r,"Glue"] \&  \mathcal{O}_\hbar(\Loc) \otimes \mathcal{O}_\hbar(\Loc)
    \end{tikzcd} 
\end{align*}

Thus the analogy of this in 3-dimension Chern-Simons theory should now be apparent from the construction in this paper: The punctures in the moduli spaces of Goncharov and Shen translate into Wilson line operators in our theory extending to infinity in the time direction. Cutting a disk around each Wilson line tangent to the boundaries gives a punctured disk two marked points on its boundary. The opposite Borel subgroups assigned to the marked points at the top and bottom can then be thought of as coming from the boundary conditions in the theory and the operation of merging two Wilson lines corresponds to the operation of gluing punctured disks together. We expect that much of the formalism described by Goncharov and Shen (the modular functor conjectures of sections 2.5  and 5 of \cite{goncharov2022quantum}) can be realized within Chern-Simons theory by exploring this connection further.

\section*{Acknowledgements}
The authors were supported by the European Research Council (ERC) under the European Union’s Horizon 2020 research and innovation programme (grant agreement No. 772960), and the Copenhagen Centre for Geometry and Topology (DNRF151).
\\Our special thanks goes to Kevin Costello, Nathalie Wahl and Ryszard Nest for helpful suggestions and discussions.

\newpage


\printbibliography

\end{document}